\RequirePackage{amsmath}
\documentclass[runningheads]{llncs}
\usepackage[T1]{fontenc}
\usepackage{multirow}
\usepackage{url}

\usepackage{amsmath}
\usepackage{hyperref}
\usepackage[misc]{ifsym}
\usepackage{orcidlink}
\usepackage{amssymb}
\usepackage{booktabs} 
\usepackage{threeparttable}
\usepackage{graphicx}
\usepackage{color}
\hypersetup{
    colorlinks=true,
    linkcolor=blue,
    filecolor=blue,
    urlcolor=blue,
    citecolor=blue
}

\begin{document}
\title{Pathological Semantics-Preserving Learning for H\&E-to-IHC Virtual Staining}
\titlerunning{Pathological Semantics-Preserving Learning} 


\author{Fuqiang Chen  %index{Chen, Fuqiang}
\inst{1,2}\orcidlink{0009-0007-2345-1348}
\and Ranran Zhang %index{Zhang, Ranran}
\inst{1}\orcidlink{0009-0005-7214-1798}
\and Boyun Zheng  %index{Zheng, Boyun}
\inst{1,2}\orcidlink{0009-0008-8645-376X}
\and Yiwen Sun %index{Sun, Yiwen}
\inst{1,3}\orcidlink{0009-0008-6973-7402}
\and Jiahui He %index{He, Jiahui}
\inst{1,4}\orcidlink{0000-0002-2152-4810}
\and Wenjian Qin  %index{Qin, Wenjian}
\inst{1(\textrm{\Letter})}\orcidlink{0000-0003-2547-0394} 
} 

\authorrunning{F. Chen et al.}

\institute{Shenzhen Institute of Advanced Technology, Chinese Academy of Sciences, Shenzhen, China \\ \email{wj.qin@siat.ac.cn}
\and Shenzhen College of Advanced Technology, University of Chinese Academy of Sciences, Shenzhen, China
\and Department of Health Technology and Informatics, The Hong Kong Polytechnic University, Hong Kong SAR, China
\and School of Computer Science, University of Nottingham Ningbo China, Ningbo, China}
\maketitle              % typeset the header of the contribution
\begin{abstract}
Conventional hematoxylin-eosin (H\&E) staining is limited to revealing cell morphology and distribution, whereas immunohistochemical (IHC) staining provides precise and specific visualization of protein activation at the molecular level. Virtual staining technology has emerged as a solution for highly efficient IHC examination, which directly transforms H\&E-stained images to IHC-stained images. However, virtual staining is challenged by the insufficient mining of pathological semantics and the spatial misalignment of pathological semantics. To address these issues, we propose the Pathological Semantics-Preserving Learning method for Virtual Staining (PSPStain), which directly incorporates the molecular-level semantic information and enhances semantics interaction despite any spatial inconsistency. Specifically, PSPStain comprises two novel learning strategies: 1) Protein-Aware Learning Strategy (PALS) with Focal Optical Density (FOD) map maintains the coherence of protein expression level, which represents molecular-level semantic information; 2) Prototype-Consistent Learning Strategy (PCLS), which enhances cross-image semantic interaction by prototypical consistency learning. We evaluate PSPStain on two public datasets using five metrics: three clinically relevant metrics and two for image quality. Extensive experiments indicate that PSPStain outperforms current state-of-the-art H\&E-to-IHC virtual staining methods and demonstrates a high pathological correlation between the staging of real and virtual stains. Code is available at \href{https://github.com/ccitachi/PSPStain}{https://github.com/ccitachi/PSPStain}.

\end{abstract}
\section{Introduction}

With the increasing demand for labor-intensive and time-consuming IHC examination \cite{anglade2020can}, virtual staining has emerged as a viable solution, which directly generates IHC-stained pathological images from conventional H\&E-stained images. This innovative approach is expected to significantly improve the efficiency of cancer diagnosis. However, the inherent challenge of virtual staining is the lack of aligned ground truth (GT) pairs for training. Generally, the virtual staining GT pair is obtained from two depth-wise consecutive cuts of the same tissue and stained separately. This inevitably prevents pixel-perfect image correspondences due to the changes in cell morphology, and staining-induced degradation \cite{li2023adaptive}.

\begin{figure}
\includegraphics[width=\textwidth]{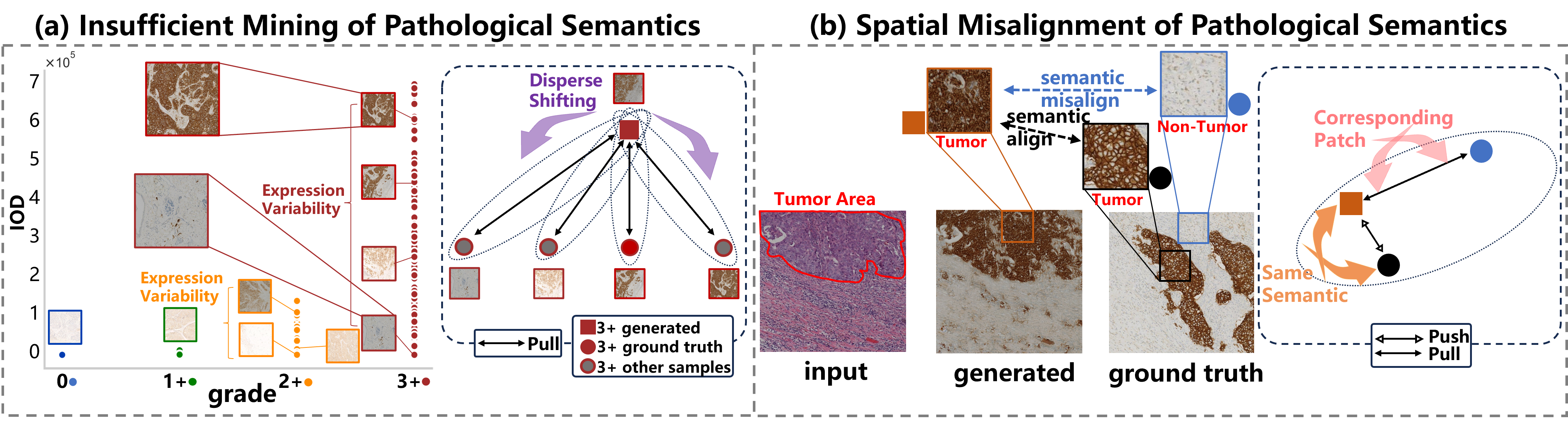}
\caption{Two key problems in H\&E-to-IHC virtual staining: \textbf{(a)} Protein expression level varies greatly in each sample from the same-grade WSI, and only preserving grade information rather than protein expression level results in a less robust representation with semantic dispersion. \textbf{(b)} Spatial misalignment between the generated image and label causes the false response of patches with similar semantics.} \label{fig1}
\end{figure}

Existing virtual staining methods can be regarded as a progressive process for extracting pathological semantic information in inconsistent GT pairs, where the most critical pathological semantics are those at the molecular level. Initially, GAN-based algorithms are directly applied in virtual staining without additional pathological constraint \cite{isola2017image,park2020contrastive,zhu2017unpaired}. Recent algorithms utilize grade-level semantic information \cite{zeng2022semi}. Subsequently, the patch-level semantic information is incorporated to enhance the molecular-level semantic preserving \cite{liu2022bci,zhang2022mvfstain,li2023adaptive}. Moreover, the pixel-level semantic information is employed with the semantic mask \cite{liu2021unpaired}.

However, rethinking the existing methods and challenges of H\&E-to-IHC virtual staining, there are two significant problems (Fig. \ref{fig1}). \textbf{(a)} \textbf{Insufficient Mining of Pathological Semantics:} The level of protein expression, which is the representation of the molecular-level semantic information, has not been directly extracted for pathological constraints. For instance, the protein expression level of human epidermal growth factor receptor-2 (HER2) varies greatly in the samples within the 3+-grade whole slide images (WSI). By only preserving grade-level pathological information rather than protein expression level for learning, the molecular-level pathological semantics become dispersed, resulting in a less robust representation.
\textbf{(b)} \textbf{Spatial Misalignment of Pathological Semantics:} The input (H\&E) and label (IHC) inevitably exhibit spatial inconsistency, implying that corresponding patches in the paired images contain different normal and tumor cells. 
Once directly assuming that similar semantics exist in corresponding patches for training, representations with similar semantics may falsely separate, and those with opposite semantics may falsely group. This interference hinders the learning of pathological representations.

To address the above problems, we propose a novel Pathological Semantics-Preserving Learning method for H\&E-to-IHC Virtual Staining (PSPStain). PSPStain comprises two novel learning strategies to solve the corresponding problems. For problem \textbf{(a)}, we directly quantify the protein expression of each IHC-stained image by determining the optical density in the DAB channel~\cite{varghese2014ihc}, preserving the molecular-level pathological semantics. Additionally, we introduce a novel Focal Optical Density (FOD) map to re-weight the contribution of tumor and non-tumor areas during quantification. For problem \textbf{(b)}, we make the assumption that the pathological content of generated image needs to be consistent with label. To enhance pathological semantic interaction despite any spatial inconsistency, we facilitate the convergence of pathological features towards cross-corresponding prototypes in the both generated image and label. During the verification process, we empirically observed that metrics such as PSNR and SSIM do not always strictly correlate with high-quality virtual staining results \cite{dubey2023structural}. This led us to incorporate three pathological relevant metrics.

The main contributions are as follows:

\textbf{1)} We propose a Protein-Aware Learning Strategy (PALS) with a Focal Optical Density (FOD) map to extract molecular-level pathological semantics, constraining the protein expression level between the generated image and label.

\textbf{2)} We propose a Prototype-Consistent Learning Strategy (PCLS), which establishes a prototypical correlation between the generated tumor and the label.

\textbf{3)} Extensive experiments on BCI and MIST-her2 datasets have demonstrated that PSPStain effectively preserves pathological semantics and improves staining performance without additional annotation.

\section{Methods}

\begin{figure}
\includegraphics[width=\textwidth]{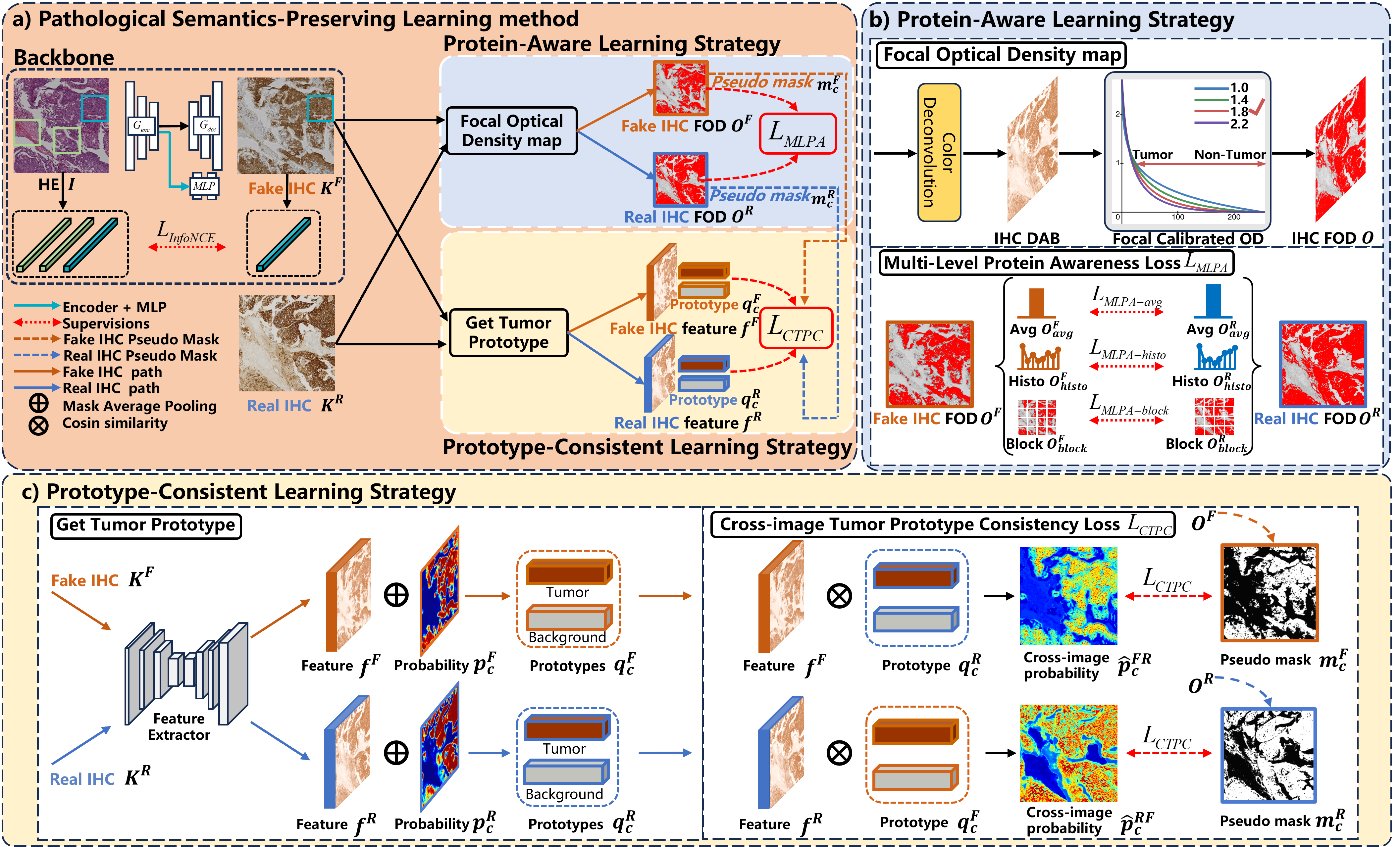}
\caption{Our proposed framework consists of two newly learning strategies for preserving protein expression consistency and enhancing semantic alignment.} \label{fig2}
\end{figure}

Our PSPStain (Fig. \ref{fig2}) contains two learning strategies: PALS, aimed at preserving the consistency of protein expression level, and PCLS, focus on enhancing tumor prototype alignment. 
In this framework, given an input pair consisting of an H\&E image $I \in\mathbb{R}^{H\times W \times C}$ and its corresponding label (real IHC) $K^R \in\mathbb{R}^{H\times W \times C}$. The backbone produces a generated image (fake IHC) $K^F \in\mathbb{R}^{H\times W \times C}$. 
% And superscript $R$ denotes the GT, superscript $F$ denotes the generated IHC-stained image.

\subsection{Protein-Aware Learning Strategy}

In most instances, protein as an antigen is conjugated to an enzyme, such as peroxidase \cite{di2012computer}.
Among the stains employed for IHC images, the DAB stain yields an intense brown staining when it reacts with the enzyme. Therefore, the fundamental concept behind PALS is to be accurately aware of molecular-level information, specifically protein expression level, by focusing on the DAB channel.

\noindent{{\textbf{Optical Density Determination:}}} Before presenting PALS, we introduce the optical density (OD). Each stain is characterized by an absorption factor to RGB. Since the OD is proportional to the concentration of the stain, the amount of stain is the factor determining the OD at a wavelength as per the LambertBeer law~\cite{varghese2014ihc}. 
\begin{equation}\label{OD}
    OD_C=-log_{10}(I_C/I_{0,C})=A \ast \boldsymbol{c}_C
\end{equation}

\noindent where $I_{0, C}$ and $I_C$ denote the light intensity entering and passing the specimen. Subscript $C$ denotes the channel. $A$ is stain amount with absorption factor $\boldsymbol{c}$. 

 \noindent{\textbf{Focal Optical Density map:}} In this module, we first use the traditional color deconvolution~\cite{ruifrok2001quantification} for stain separation. Then we specifically select DAB stain’s OD values to generate the RGB image (IHC DAB). 

 In most samples, the tumor area is significantly smaller than the non-tumor area, resulting in a notable imbalance. Hence, training with just the normal OD map Eq.\ref{OD} becomes inefficient as the majority of locations represent easy non-tumor areas, contributing no useful learning signal. Instead, we propose to reshape the map function to down-weight non-tumor area and meanwhile focus training on the tumor area. For FOD, we simulate Eq.\ref{OD} by converting the IHC DAB to grayscale and using the focal calibrated map to assign gray values to positive signal (FOD) as followed:
\begin{equation}
    {O_C}=(-log_{10}((I_C)/I_{0,C}))^{\alpha}
\end{equation}

\noindent where $O$ is the FOD with tunable focusing
 parameter $\alpha > 1$.
 The FOD map is visualized for several values of $\alpha$ in Fig.\ref{fig2}(b). With an increase in the focusing parameter $\alpha$, the de-emphasis of non-tumor areas and the emphasis on tumor areas are enhanced. Among the tested values, $\alpha=1.8$ yields the best results.

\noindent{\textbf{Multi-Level Protein Awareness:}} For pathology consistency, a Multi-Level Protein Awareness (MLPA) loss is adopted to match fake IHC FOD $O^F$ with real IHC FOD $O^R$.

 Initially, we align the average protein expression level and set a tolerance value $\beta$ of 0.2 to account for inherent differences. If the difference in protein expression level is less than $\beta$ multiplied by the average expression level in reference image $O_{avg}^R$, there is no contribution to model learning.
\begin{equation}
L_{MLPA-{avg}} =  
\begin{cases}
\left\|O_{avg}^F-O_{avg}^R\right\|_{2} ,& if |O_{avg}^F-O_{avg}^R| \geq \beta \cdot O_{avg}^R \\
0,&if |O_{avg}^F-O_{avg}^R|<\beta \cdot O_{avg}^R
\end{cases}
\end{equation}

 \noindent The $O_{avg}$ denotes the average amount of IHC FOD $O$, $O_{avg}\!=\!\frac{1}{H\!\times\! W} \!\sum_{h=1}^{H}\!\sum_{w=1}^{W}\!O_{h,w}$.

 Secondly, we divide the FOD range from 0 to $e$ into intervals of $N_h=20$ to ensure consistency in the distribution of protein expression between generated images and labels. 
 \begin{equation}
     L_{MLPA-{histo}} = \frac{1}{N_h}\sum_{i=1}^{N_h} \left\|O_{histo_i}^F-O_{histo_i}^R\right\|_{2}
 \end{equation}
 \noindent where $O_{histo_i}$ denotes the $i_{th}$ histogram accumulation of the IHC FOD $O$. $O_{histo_i} = \sum_{j=1}^{M_h} O_j$,$if \frac{(i-1) \cdot e}{N_h}<O_j\leq\frac{(i) \cdot e}{N_h}$, where $O_j$ denotes the $j_{th}$ pixel of IHC FOD $O$ and $M_h$ is the number of pixels in the $i_{th}$ histogram.

 Finally, we divide the image into $N_b=16$ blocks and independently calculate the average expression amount of each block. This preserves the intensity of expression in specific regions as regularization, referred to as $MLPA-block$.
 \begin{equation}
     L_{MLPA-{block}} = \frac{1}{N_b}\sum_{i=1}^{N_b} \left\|O_{block_i}^F-O_{block_i}^R\right\|_{2}
 \end{equation}
\noindent where $O_{block_i}$ denotes the $i_{th}$ block average amount of the IHC FOD $O$. $O_{block_i} = \frac{1}{M_b}\sum_{j=1}^{M_b} O_j, if\ position(O_j)\in block_i $ 
,where $O_j$ denotes the $j_{th}$ pixel of IHC FOD $O$ and $M_b$ is the number of pixels in the $i_{th}$ block.

Thus, the overall objective of our MLPA is formulated as follows:
\begin{equation}
    L_{MLPA} = L_{MLPA-{avg}} + L_{MLPA-{histo}} + L_{MLPA-{block}}
\end{equation}

\subsection{Prototype-Consistent Learning Strategy}
 Inspired by \cite{zhang2023self}, PCLS effectively preserves pathological semantic consistency even in the presence of spatial misalignment. Specifically, when comparing the generated image $K^F$ with the label $K^R$, it is highly likely that the tumor content within these images remains consistent. Therefore, we calculate the prototype consistency loss $L_{CTPC}$ between them. We derive masks $m_c^F$ and $m_c^R$ by applying the same threshold to fake IHC FOD $O^F$ and real IHC FOD $O^R$.

\noindent{\textbf{Get Tumor Prototype:}}
By defining the tumor class and non-tumor class in the IHC image, the prototype of each class refers to the aggregated representation. We pretrain a segmentation UNet with the pseudo mask $m_c^R$ on real IHC images and freeze it to extract tumor semantic. Initially, input IHC image is processed by UNet, yielding a seg probability map as output. Feature maps are then derived from the layer preceding the output layer.
Let $f^F \in \mathbb{R}^{H\times W \times D}$ represents the feature map of output $K^F$, while $p_c^F(i)$ denotes the probability of pixel $i$ belonging to class $c$. Similarly, $f^R \in \mathbb{R}^{H\times W \times D}$ represents the feature map of label $K^R$, and $p_c^R(i)$ denotes the probability of pixel $i$ belonging to class $c$. Using these feature maps and probability maps, we aggregate class-wise prototypes representing pixel-wise features of tumor and background. The prototypes of output $q_c^F$ and label $q_c^R$ are as followed:
\begin{align}
    {q}_c^{{F}}=\frac{\sum_i p_c^F(i)\cdot f^F(i)}{\sum_i p_c^F(i)} ,
    & {q}_c^{{R}}=\frac{\sum_i p_c^R(i)\cdot f^R(i)}{\sum_i p_c^R(i)} 
\end{align}

 \noindent{{\textbf{Cross-image Tumor Prototype Consistency:}}} 
The feature similarity is calculated based on the output prototype $q_c^F$ and the label prototype $q_c^R$. For semantic interaction, cosine similarity is computed leveraging image features with prototypes from another image. Specifically, $\hat{s}_{c}^{FR}$ and $\hat{s}_{c}^{RF}$ are two cross-image prototypical similarity maps obtained by calculating cosine similarity between the feature map $f^F$, $f^R$ and the prototype vector $q_c^R$, $q_c^F$. These are defined as:
%\boldsymbol
\begin{align}
    \hat{s}_{c}^{FR}=\frac{f^F\cdot{q}_c^R}{\|f^F\|\cdot\left\|{q}_c^R\right\|},
    & \hat{s}_{c}^{RF}=\frac{f^R\cdot{q}_c^F}{\|f^R\|\cdot\left\|{q}_c^F\right\|}
\end{align}

Then, we apply softmax operations to $\hat{s}_c^{FR}$ and $\hat{s}_c^{RF}$ to derive the corresponding cross-image probability prediction $\hat{p}_c^{FR}$ and $\hat{p}_c^{RF}$ respectively:
%\in \mathbb{R}^{H\times W \times C}
\begin{align}
    \hat{p}_c^{FR} =\frac{e^{\hat{s}_{c}^{FR}}}{\sum_{c}e^{\hat{s}_{c}^{FR}}},
    & \hat{p}_c^{RF} = \frac{e^{\hat{s}_{c}^{RF}}}{\sum_{c}e^{\hat{s}_{c}^{RF}}}
\end{align}

The cross-image tumor prototype consistency (CTPC) loss is defined as:
 \begin{equation}
     L_{CTPC}=
     \begin{aligned}
     \frac{1}{C\!\times H\!\times W}\!\sum_{i=1}^{H\times W}\!\sum_{c=1}^{C}
     (\left\| \hat{p}_c^{FR}(i)-m_c^F(i)\right\|_{2}\!+\!
     \left\|\hat{p}_c^{RF}(i)-m_c^R(i)\right\|_{2})
     \end{aligned}
 \end{equation}
 
\subsection{Loss Function of PSPStain.}

The overall learning objective is as follows:
\begin{equation}
    L_{total} = L_{adv} + L_{NCE} + \lambda_{M}L_{MLPA} +  \lambda_{C}L_{CTPC} + \lambda_{S}L_{SSIM} + \lambda_{G}L_{GP}
\end{equation}
where $L_{MLPA}$, $L_{CTPC}$, $L_{GP}$ focus on pathological consistency, with $L_{GP}$ originating from \cite{liu2022bci}. $L_{adv}, L_{NCE}$, $L_{SSIM}$ contribute to image quality enhancement.

\section{Experiments}

\noindent{{\textbf{Datasets:}}}
Experiments are conducted on two public datasets: the Breast Cancer Immunohistochemical (BCI) challenge dataset~\cite{liu2022bci} and the Multi-IHC Stain Translation (MIST) dataset \cite{li2023adaptive}. The BCI dataset consists of 3396 H\&E-HER2 pairs of the training set images and 977 pairs of test set images from 51 WSIs. In the MIST dataset, we use the ${\rm MIST_{HER2}}$, which contains 4642 paired samples for training and 1000 for testing from 64 WSIs.% The expression levels of HER2 in these 51 patients cover four grades of 0, 1+, 2+ and 3+.

\noindent{{\textbf{Implementation Details:}}} CUT~\cite{park2020contrastive} is selected as the baseline model. The generator is ResNet-6Blocks~\cite{he2016deep} and the discriminator is PatchGAN~\cite{isola2017image}. We trained our networks with random $512\times512$ crops and a batch size of four. Adam optimizer with the learning rate of $1\times10^{-4}$. The maximum number of training epochs was set to 80. The weight valule of $\lambda_{M},\lambda_{C},\lambda_{S},\lambda_{G}$ is 1.0, 2.5, 0.05, 10.0.

\renewcommand{\arraystretch}{1.1} %控制行高
\begin{table}[h]
 
  \centering
  \setlength\tabcolsep{3pt} 

  \fontsize{9}{10}\selectfont
  
  \caption{Quantitative evaluations on two datasets using three pathology-related metrics and two image quality metrics. The mIOD and IOD are subtracted from the GT and the value closer to 0 indicates better results. The \textbf{best} values are highlighted.}
  \label{tab:performance_comparison}

  \resizebox{\linewidth}{!}{
    \begin{tabular}{c|ccccc|ccccc}
    % \toprule
    \hline
    \multirow{3}{*}{Method}&
    \multicolumn{5}{c|}{ BCI-Her2}&\multicolumn{5}{c}{ MIST-Her2}\cr
    \cline{2-11}
    &\multicolumn{3}{c|}{Pathological Relevance}&\multicolumn{2}{c}{ Image Quality }&\multicolumn{3}{|c|}{Pathological Relevance }&\multicolumn{2}{c}{ Image Quality }\cr
    \cline{2-11}
    % \cmidrule(lr){2-4} \cmidrule(lr){5-7}
    &mIOD&IOD$_{\times 10^{7}}$&Peason-R$\uparrow$&PSNR$\uparrow$&SSIM$\uparrow$&mIOD&IOD$_{\times 10^{7}}$&Peason-R$\uparrow$&PSNR$\uparrow$&SSIM$\uparrow$\cr
    \hline
    % \midrule
    CUT~\cite{park2020contrastive}&-0.3528&-2.9126 &0.0276 &18.1246&0.4483&-0.0830&-2.9478&0.7164&13.8998&0.1680\cr
    Pix2Pix~\cite{isola2017image}&-0.3598&-2.8819  &0.0431&15.8659&0.4263&\textbf{-0.0414}&-2.5782&0.2986&12.9840&0.1676\cr
    PyramidP2P~\cite{liu2022bci}&-0.3105&-2.9056 &0.1018&\textbf{19.9488}&0.4647&-0.0446&-4.5285&0.6894&\textbf{14.9122}&0.1995\cr
    ASP~\cite{li2023adaptive}&-0.2438&-2.9292 &-0.0406&17.8651&\textbf{0.4923}&-0.1303&-5.7422&0.0659&14.1841&\textbf{0.2004}\cr
    \hline
    
    {\bf PSPStain}&\textbf{-0.1800}&\textbf{-0.7808} &\textbf{0.7553}&18.6220&0.4498&-0.0834&\textbf{-2.5491}&\textbf{0.8303}&14.1948&0.1876\cr
     \hline
    
    \end{tabular}
    }
    
\end{table}

\begin{figure}[hb]
\centering
\includegraphics[width=\textwidth]{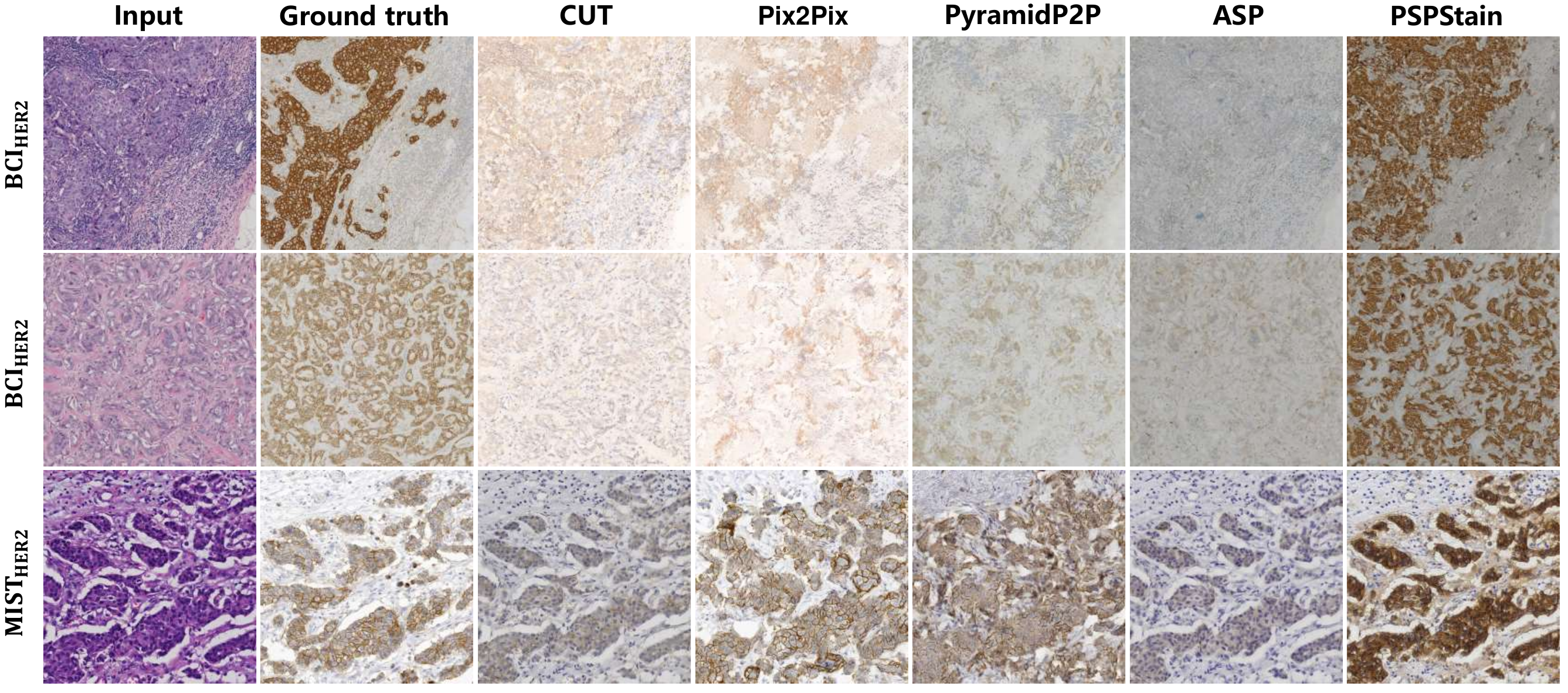}
\caption{Qualitative comparison with samples virtually stained by various methods.} \label{fig_visualization}
\end{figure}

\noindent{{\textbf{Evaluations:}}} To evaluate image quality, we use the standard Peak Signal-to-Noise Ratio (PSNR
) \cite{avcibas2002statistical} and Structural Similarity Index Measure (SSIM) \cite{wang2004image}. To explore whether virtual staining meets clinical needs, we evaluate the positive signal prediction using integrated optical density (IOD) and mean integrated optical density (mIOD)~\cite{wang2009survivin,zhang2022mvfstain}, and calculate the Pearson correlation coefficient (Pearson-R)~\cite{liu2021unpaired}. Specifically, IOD quantifies the amount of positive signal statistically, while mIOD represents the intensity of positive signal within the positive area. Pearson-R is used to assess the pathological correlation with protein expression level rather than just the size of the positive area as discussed in \cite{liu2021unpaired}.

\begin{figure}[t]
\centering
\includegraphics[width=\textwidth]{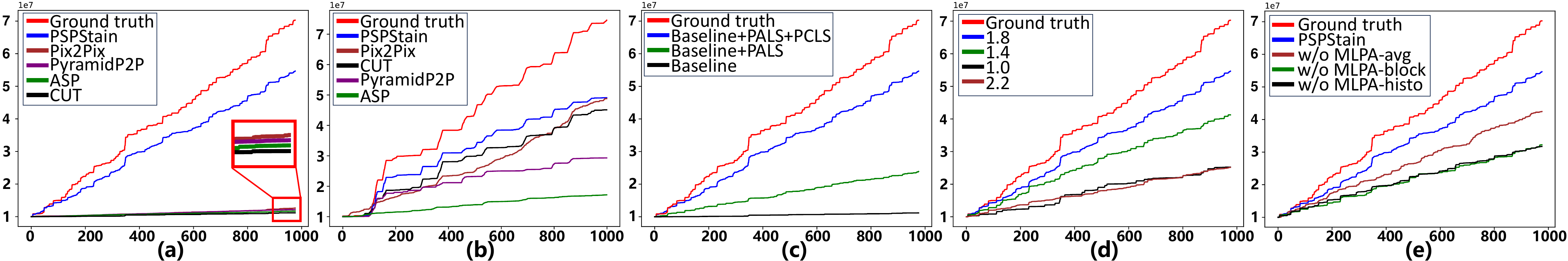}
\caption{The abscissa represents the index number of samples. The ordinate is the accumulated IOD value. \textbf{(a) (b)} refer to the curve on BCI and MIST datasets, respectively. \textbf{(c)} Strategy ablation. \textbf{(d)} FOD factor $\alpha$ ablation. \textbf{(e)} MLPA module ablation}. \label{fig_ablation}
\end{figure}

\noindent{{\textbf{Comparison Study:}}}
We compare the performance with state-of-the-art methods. Empirical results demonstrate the superiority of our method.
Table. \ref{tab:performance_comparison} shows that PSPStain achieves the lowest deviation of mIOD and IOD and the highest Pearson-R score from the labels, indicating the highest pathological consistency. The higher PSNR and SSIM indicate higher image quality however not always in virtual staining.
Specifically, the blurry IHC images with more averaged pixel value may cause them to be increased within spatial misalignment~\cite{zhu2023breast}.
% Because PSNR is calculated on a pixel-by-pixel basis, and SSIM, by default, involves sliding a 7x7 window for computation. Hence, they can be affected by spatial misalignment. Specifically, the blurry IHC images with more averaged pixel value may cause them to be inflated~\cite{zhu2023breast}.
Fig. \ref{fig_visualization} shows the qualitative results, indicating that PSPStain effectively and robustly highlights tumor regions.
Fig. \ref{fig_ablation}\textbf{(a)(b)} compares the performance of different methods in two datasets by accumulating positive signals in each sample, revealing that PSPStain closely matches the ensemble of GT more than others.

%%%%%%%%%%%%%%%%%%%%%%%%%%%%%%%%%%%%%%%%%%%
\renewcommand{\arraystretch}{1.0} %控制行高
\begin{table}[tp]
 
  \centering
  \setlength\tabcolsep{3pt} 

  \fontsize{9}{10}\selectfont
  
  \caption{Ablation studies on Strategy, FOD factor, and MLPA module. All ablation study is performed on the BCI dataset. The \textbf{best} values are highlighted.}
  \label{tab:performance_ablation}
  \resizebox{\linewidth}{!}{

    \begin{tabular}{cc|ccccc}
    % \toprule
    \hline
    \multicolumn{2}{c|}{}                                    & \multicolumn{5}{c}{BCI-Her2}                        \\ \cline{3-7} 
\multicolumn{2}{c|}{}                                             & \multicolumn{3}{c|}{Pathological Relevance} & \multicolumn{2}{c}{Image Quality} \\ \cline{3-7} 
\multicolumn{2}{c|}{\multirow{-3}{*}{Method}}                            &mIOD&IOD$_{\times 10^{7}}$&Peason-R$\uparrow$&PSNR$\uparrow$&SSIM$\uparrow$         \\ \hline
\multicolumn{1}{c|}{}                   & Baseline  &-0.3524&-2.9551 &0.3611 &\textbf{21.1427}&\textbf{0.5043}         \\
\multicolumn{1}{c|}{}                   & Baseline+PALS&  -0.2894&-2.3197&0.6197&19.7659&0.4826          \\
\multicolumn{1}{c|}{\multirow{-3}{*}{Strategy ablation}} &  Baseline+PALS+PCLS &\textbf{-0.1800}&\textbf{-0.7808} &\textbf{0.7553}&18.6220&0.4498          \\ \hline
\multicolumn{1}{c|}{}                   & 1.0  &-0.2108&-2.2525 &0.5884 &\textbf{19.5469}&0.4419         \\ 
\multicolumn{1}{c|}{}                   & 1.4  &-0.2091&-1.4506 &0.7223 &19.3400&0.4366         \\ 
\multicolumn{1}{c|}{}                   & 1.8  &\textbf{-0.1800}&\textbf{-0.7808} &\textbf{0.7553}&18.6220&0.4498          \\
\multicolumn{1}{c|}{\multirow{-4}{*}{FOD factor $\alpha$ ablation}} & 2.2 &-0.2516&-2.2563 &0.5630&19.5217&\textbf{0.4656}          \\ \hline
\multicolumn{1}{c|}{}                   & w/o MLPA-avg  &\textbf{-0.0687}&-1.3915 &0.0474 &19.0137&0.4353         \\ 
\multicolumn{1}{c|}{}                   & w/o MLPA-histo &-0.2286&-1.9272 &0.6972 &\textbf{19.1779}&0.4338         \\ 
\multicolumn{1}{c|}{}                   & w/o MLPA-block  &  -0.1616&-1.9022&0.6728&17.0345&0.4080          \\
\multicolumn{1}{c|}{\multirow{-4}{*}{MLPA module ablation}} & PSPStain &-0.1800&\textbf{-0.7808} &\textbf{0.7553}&18.6220&\textbf{0.4498}          \\ \hline

    \end{tabular}
    }
  
\end{table}

\noindent{{\textbf{Ablation Study:}}}
As demonstrated in Table. \ref{tab:performance_ablation} and Fig. \ref{fig_ablation}\textbf{(c)(d)(e)}, we conduct the ablation study using the BCI dataset. \textbf{(c)} For strategy ablation, both learning strategies effectively improve the performance of virtual staining. \textbf{(d)} For FOD ablation, we test various values of $\alpha$. When $\alpha$ is set to 1.8, PSPStain performs the best. \textbf{(e) }For MLPA ablation, all three modules prove to be beneficial.
%All of these values yielded beneficial results.
%we conduct the ablation using the BCI dataset

\section{Conclusion}
In this paper, we propose PSPStain which aims to extract molecular-level pathological information and align cross-image pathological semantics despite any spatial inconsistency. Specifically, protein expression level as the representation of molecular-level semantics is preserved by PALS. The FOD map, integrated into PALS, focuses training on the tumor rather than the non-tumor area. PCLS maintains semantic alignment via cross-image prototypical consistency learning.
The superior virtual staining performance shows that PSPStain effectively and robustly preserves pathological semantics without additional expert annotation.
Our method has been validated on the Her2-image dataset and can be extended to other DAB-stained IHC images.
% Our method is designed for all DAB-stained IHC images, and we have tested it on the Her2 image. Additional IHC datasets will be tested in future work.

\begin{credits}
\subsubsection{\ackname} This work was supported by the National Natural Science Foundation of China (No. 62271475), Ministry of Science and Technology's key research and development program (2023YFF0723400), the Youth Innovation Promotion Association CAS (2022365) and Shenzhen-Hong Kong Joint Lab on Intelligence Computational Analysis for Tumor lmaging (E3G111).

\subsubsection{\discintname}
The authors declare no competing interests in the paper.

\end{credits}
%(optional) acknowledgments\footnote{If EquinOCS, our proceedings submission
% system, is used, then the disclaimer can be provided directly in the system.},
% for example: The authors have no competing interests to declare that are
% relevant to the content of this article. Or: Author A has received research
% grants from Company W. Author B has received a speaker honorarium from
% Company X and owns stock in Company Y. Author C is a member of committee Z.
\bibliographystyle{splncs04}
\bibliography{Paper-2078}

\end{document}